\documentclass[journal]{IEEEtran}

\ifCLASSINFOpdf
\usepackage[pdftex]{graphicx}
\graphicspath{{../pdf/}{../jpeg/}{figs/}}
\else
\usepackage[dvips]{graphicx}
\graphicspath{{../eps/}{figs/}}
\DeclareGraphicsExtensions{.eps}
\fi

\usepackage{bm, amsmath, amssymb, color, amsfonts, amsthm, array, threeparttable, multirow, stfloats, subfigure, colortbl, pdflscape, cite, url, CJK}

\usepackage[switch]{lineno}

\usepackage{algpseudocode}
\usepackage[ruled,linesnumbered,vlined]{algorithm2e}

\usepackage[table*]{xcolor}
\xdefinecolor{graylt}{gray}{0.85}
\xdefinecolor{graydk}{gray}{0.65}

\theoremstyle{plain}

\theoremstyle{definition}

\theoremstyle{remark}

\begin{document}


\begin{CJK*}{UTF8}{gkai}

\title{Multiobjective Multitasking Optimization Based on Decomposition with Dual Neighborhoods}

\author{Xianpeng~Wang,~\IEEEmembership{Member,~IEEE},~Zhiming~Dong, \\ ~Lixin~Tang,~\IEEEmembership{Senior Member,~IEEE} and Qingfu Zhang~\IEEEmembership{Fellow,~IEEE}, 
	\thanks{This research was supported by the National Key Research and Development Program of China (2018YFB1700404), the Fund for the National Natural Science Foundation of China (62073067), the Major Program of National Natural Science Foundation of China (71790614), the Major International Joint Research Project of the National Natural Science Foundation of China (71520107004), and the 111 Project (B16009). (Corresponding author: Lixin Tang)} 
	\thanks{X. Wang is with the Key Laboratory of Data Analytics and Optimization for Smart Industry (Northeastern University), Ministry of Education, Shenyang, 110819, China (e-mail: wangxianpeng@ise.neu.edu.cn).}
	\thanks{Z. Dong is with the Liaoning Engineering Laboratory of operation Analytics and Optimization for Smart Industry, Liaoning Key Laboratory of Manufacturing System and Logistics, Shenyang, 110819, China (e-mail: dongzm@stumail.neu.edu.cn).}
	\thanks{L. Tang is with the Institute of Industrial and Systems Engineering, Northeastern University, Shenyang, 110819, China (e-mail: qhjytlx@mail.neu.edu.cn).}
	\thanks{Q. Zhang is with the City University of Hong Kong, Shenzhen Research Institute, Shenzhen 518057, China (qingfu.zhang@cityu.edu.hk).}
}
	
\markboth{Journal of \LaTeX\ Class Files,~Vol.~x, No.~x, x~xxxx}%
{Shell \MakeLowercase{\textit{et al.}}: Bare Demo of IEEEtran.cls for IEEE Journals}

\maketitle

\begin{abstract}
	This paper proposes a multiobjective multitasking optimization evolutionary algorithm based on decomposition with dual neighborhood. In our proposed algorithm, each subproblem 	
	 not only maintains a neighborhood based on the Euclidean distance among weight vectors within its own task, but also keeps a neighborhood with subproblems of other tasks. Gray relation analysis is used to define neighborhood among subproblems of different tasks. In such a way, relationship among different subproblems can be effectively exploited to guide the search. 
	Experimental results show that our proposed algorithm outperforms four state-of-the-art multiobjective multitasking evolutionary algorithms and a traditional decomposition-based multiobjective evolutionary algorithm on a set of test problems. 
\end{abstract}

\begin{IEEEkeywords}
	Multiobjective multitasking optimization, evolutionary algorithm, decomposition, grey relation analysis.
\end{IEEEkeywords}

%
\IEEEpeerreviewmaketitle

\section{Introduction} \label{s1}


\IEEEPARstart{M}{ultitasking} evolutionary optimization \cite{Gupta2016} is a new growing research area. Borrowing the idea from multitasking learning, multitasking optimization (MTO) explores relationship among different tasks for improving the search efficiency, and it can also distinguish and make use of differences among these tasks.
 
Evolutionary algorithms (EAs) are widely used for solving optimization problems~\cite{Wang2018, tang2020data}. Multitasking evolutionary algorithms (MTEAs) \cite{Gupta2016, Gupta2017} help the optimization of different tasks by sharing a same population and mining the implicit information among tasks. On the one hand, such sharing and mining strategies can speed up the optimization process of different tasks. On the other hand, it can help each individual optimization process escape its local optima through the interaction between tasks. Typically, for bi-level optimization problems \cite{Huang2020}, when multiple upper level candidate solutions are analyzed simultaneously, the lower level optimization problem can be considered as an MTO problem \cite{Gupta2015}.
In the field of expensive optimization \cite{Ding2019},  knowledge of computationally cheap optimization problems is transferred to expensive optimization problem through multitasking evolutionary algorithm to improve the convergence speed of expensive problem.
Yi \emph{et al.} \cite{Yi2020} transformed the problem with interval uncertainty into an MTO problem.
Feng \emph{et al.} \cite{FengZhou2019} used a multitasking evolutionary algorithm to deal with a generalized variant of vehicle routing problem with occasional drivers to cope with the requirement that multiple tasks in cloud computing services should be optimized at the same time.
In addition, MTEAs have been studied and employed to successfully solve different problems such as the optimization of operational indices in beneficiation processes \cite{Yang2019}, composition of cloud computing service \cite{Bao2018}, sparse reconstruction \cite{Li2019}, bi-fidelity optimization \cite{Wang2020}, hyper-heuristics \cite{Hao2020}, and multiobjective pollution-routing problem \cite{Rauniyar2019}.

Based on NSGA-II \cite{Deb2002}, Gupta \emph{et al.} \cite{Gupta2017} proposed an evolutionary algorithm for solving multiobjective multitasking optimization problems, called MO-MFEA, and used it to solve two composites manufacturing problems (two multiobjective optimization problems, i.e., MOPs).

Decomposition-based multiobjective evolutionary algorithm (MOEA/D) has been widely used in multiobjective optimization~\cite{Zhang2007, Li2009}. Yao \emph{et al.} \cite{Yao2020} proposed a decomposition based algorithm for solving multiobjective multitasking optimization problem. Their algorithm does not make use of relationship among different tasks very well. 
Neighborhood structure is used for establishing relationship among different subproblems in MOEA/D algorithms. It is assumed that subproblems with close Euclidean distances between weight vectors have similar optimal solutions \cite{Zhou2016} in MOEA/D. However,  relationship between subproblems of different tasks cannot be measured by weight vectors because they belong to different tasks.
To efficiently mine and use the relationship between subproblems of different tasks, this paper proposes a multiobjective multitasking optimization evolutionary algorithm based on decomposition with dual neighborhood, denoted as MTEA/D-DN. It defines and uses a neighborhood structure based on the Euclidean distance between the weight vectors within its own task, denoted as internal neighborhood, each subproblem also has an external neighborhood relationship with the subproblems of other tasks defined by grey relation analysis \cite{Liang1999}. During the evolution of the population, the transfer of information between different tasks is achieved  by exchanging information between these two neighborhoods.

The remaining part of the paper proceeds as follows. Section \ref{s2} introduces some basic concepts and related work. The proposed multiobjective multitasking evolutionary algorithm based on decomposition with dual neighborhood is described in details in Section \ref{s3}. Then, some computational experiments and discussion are presented in Section \ref{s4}. Finally, the paper is concluded in Section \ref{s5}.

\section{Background and Related Work} \label{s2}

In this section, we first describe some basic definitions of the multiobjective MTO. Then, the decomposition strategy and the grey relation analysis are briefly introduced. Finally, a review of currently existing evolutionary algorithms for multiobjective MTO is presented.

\subsection{Multiobjective Multitasking Optimization}

In general, a multiobjective multitasking optimization problem that minimizes all objectives of each task can be defined as follows \cite{Gupta2017}: 
\begin{eqnarray}
	\label{momfo}
	\begin{aligned}
		& \min~\mathcal{F}(\mathbf{x}_1, \mathbf{x}_2, \ldots, \mathbf{x}_K)  \\
		& ~~~~~ = \min~(\mathbf{F}_1(\mathbf{x}_1),\mathbf{F}_2(\mathbf{x}_2),...,\mathbf{F}_K(\mathbf{x}_K))  \\
		& {\rm s.t.}~\mathbf{x}_i \in \mathbf{\Omega}_i,~i \in \{1,2,\ldots,K\}
	\end{aligned}
\end{eqnarray}
where, $\mathbf{F}_i$ is the $i$-th task to be optimized (an MOP), $\mathbf{x}_i = (x^1_i, x^2_i, \ldots, x^{D_i}_i)$ is the $D_i$-dimensional decision variable vector, and $\mathbf{\Omega}_i$ denotes the feasible domain of task $i$.
Here, for each task, there are multiple objectives that need to be optimized at the same time, and there are conflicting relationship between these objectives.
The goal is to find a representative Pareto front (PF) for each multiobjective optimization task, so as to help decision makers analyze the relationship between different objectives and then make reasonable trade-offs and decisions \cite{Dong2020}.
Optimizing each task individually is the most straightforward approach.
However, in similar environments, there may be implicit relationship between these tasks, so it is necessary to explore and exploit potentially useful information between them to improve the efficiency of task solving.

\subsection{Decomposition Strategy}

A series of weight vectors and a scalarizing function \cite{Pescador2017} are two components of the decomposition strategy in decomposition-based multiobjective evolutionary algorithms.
Each weight vector and multiple objectives based on the scalarizing function constitute a single-objective optimization subproblem, and then a population is employed to optimize these subproblems simultaneously \cite{Zhang2007,Li2009}.
The Euclidean distance between the weight vectors defines the neighborhood structure between these subproblems.
It is generally assumed that subproblems with closer weight vectors in the same multiobjective optimization problem have similar optimal solutions \cite{Zhou2016}. 
Therefore, during population evolution, the exploitation and exploration of the algorithm are balanced by the neighborhood structure (whether the selection and update of parents comes from the neighborhood).
Based on the idea of decomposition, some variants of decomposition-based evolutionary algorithm have been proposed, such as embedded dynamical resource allocation \cite{Zhang2009,Zhou2016,DongWang2020}, angle-based decomposition \cite{Cheng2016}, decomposing a multiobjective problem into some simple MOPs \cite{Liu2014,Gu2018}, combining domination-based strategy \cite{Deb2014,Li2015,Tang2019}, etc..

\subsection{Grey Relation Analysis}

Grey relation analysis quantifies the degree of similarity or dissimilarity between different factors by calculating the numerical relationship between them, i.e., reference sequence and a number of compared sequences, in order to evaluate whether the factors are closely related or not.
For a given normalized reference sequence $Y = \{y(k) \mid k = 1, 2, \ldots, n\}$ and the compared sequence $X_i = \{x_i(k) \mid k = 1, 2, \ldots, n\}$, $i = 1, 2, \ldots, m$, the grey relational degree of the reference sequence to a compared sequence can be calculated in the following form \cite{Liang1999}: 
\begin{eqnarray}
	r_i = \frac{1}{n} \sum_{k=1}^{n} \frac{\min\limits_{i=1}^{m} \min\limits_{k=1}^{n} \Delta_i(k) + \rho \times \max\limits_{i=1}^{m} \max\limits_{k=1}^{n} \Delta_i(k)}{\Delta_i(k) + \rho \times \max\limits_{i=1}^{m} \max\limits_{k=1}^{n} \Delta_i(k)}
	\label{gra}
\end{eqnarray}
where $\Delta_i(k) = |y(k) - x_i(k)|$, $n$ is the dimension of each factor, and $m$ is the number of compared sequences. $\rho \in [0, 1]$ is the distinguished coefficient, the smaller the value of $\rho$, the greater its distinguished ability will be, and its value is usually set as $\rho=0.5$ \cite{Liang1999}.

\subsection{Related Work}

The basic scheme of MTEA is to map the decision space of all tasks into a \emph{unified search space}, and to optimize these tasks simultaneously with a single population that is encoded in the \emph{unified search space} \cite{Gupta2016, Gupta2017}. 
Furthermore, information transfer between tasks is performed by updating the current population with offspring that is generated by the genetic operator, with a certain probability (i.e., random mating probability \emph{rmp}), from individuals associated with different tasks. 
Such information transfer is also known as implicit information transfer \cite{Feng2019}.
Feng \emph{et al.} \cite{Feng2019} argued that the method of knowledge transfer through genetic operator in \cite{Gupta2016,Gupta2017} limits the use of other evolutionary search operators so that some high-performance evolutionary search operators cannot be embedded in existing multitasking evolutionary algorithms.
Further, Feng \emph{et al.} \cite{Feng2019} established the mapping relationship of different tasks in the decision space through autoencoder technology to transfer knowledge, which is called explicit transfer.
It should be noted that training an autoencoder network is a time-consuming process.
Tang \emph{et al.} \cite{Tang2020} applied principle component analysis method to map the domains of multiple tasks to low-dimensional aligned subspace, and employed this subspace for information transfer.
In addition, Feng \emph{et al.} \cite{Feng2020} tackled the problem of knowledge transfer between tasks by constructing weighted l$_1$-norm-regularized reconstruction error between different combinatorial optimization problems.
In order to harness the unique performance of different crossover operators, Zhou \emph{et al.} \cite{Zhou2020} proposed an adaptive knowledge transfer strategy based on multiple crossover operators.
Yao \emph{et al.} \cite{Yao2020} proposed a multiobjective multitasking evolutionary algorithm based on decomposition strategy \cite{Zhang2007}, in which the way of information transfer between tasks can be summarized as: the offspring generated by the parents associated with the same task can be used to update the individuals associated with other tasks. However, this information transfer method did not explore and mine the relationship between tasks.

On the other hand, Lin \emph{et al.} \cite{Lin2019} used incremental Naive Bayes classifier to select several individuals from other task to the target task as transfer knowledge to participate in the production of offspring, and then used these selected individuals as training data to update the classification model.
Bali \emph{et al.} \cite{Bali2020_1,Bali2020_2} pointed out that negative information transfer may impair the solution of the task. Furthermore, they employed data-driven technology to analyze the overlaps in the probabilistic search distributions between the different tasks and adjusted the probability of information transfer between tasks to prevent negative transfer.
Similarly, Zheng \emph{et al.} \cite{Zheng2020} suggested that the fixed probability of information transfer limits the sharing and utilization of useful knowledge between tasks, and proposed a self-regulated strategy based on the \emph{ability vector}.

\section{Proposed Algorithm} \label{s3}
In this section, we first give a detailed description of our proposed algorithm, and then some discussion of the proposed algorithm are presented.

\subsection{Algorithm Framework} \label{s31}

The framework of our proposed algorithm is shown in Algorithm \ref{algo framework}.
At the beginning, the initialization of the items such as the internal and external neighborhoods structure $B$ and $\widetilde{B}$, and the task index $\Phi$ in the proposed algorithm are carried out, as shown in line 1, which is explained in detail in Algorithm \ref{initialization}.
In the main loop of the algorithm, the joint set $\mathcal{U}$ of each sub-population $\mathcal{P}^{i}$ that corresponds to task $i$ after shuffling is firstly traversed, as shown in line 3.
Then, for each individual $\mathbf{x}$ in the joint set $\mathcal{U}$, the following procedure will be performed step by step. First, the index of the task for which the current individual $\mathbf{x}$ belongs to and the subproblem index that $\mathbf{x}$ is matched are obtained respectively (lines 5 and 6). 
Secondly, the candidate set $Q$, i.e., a set of subproblem indexes that is used to generate the offspring, is determined (line 7).
It should be noted that the determination of $Q$ is preceded by determining which task the candidate set $Q$ comes from (please refer to Algorithm \ref{neighborhood selection} for details).
The next step is the production of offspring, which is introduced in Algorithm \ref{reproduction}.
Finally, the generated offspring is used to update the current state of some items, whose detailed steps are shown in Algorithm \ref{update}.
In the following, we further elaborate on Algorithms \ref{initialization}-\ref{update}, respectively.

\begin{algorithm}
	\caption{Framework of MTEA/D-DN}
	\label{algo framework}
	\KwIn{sub-population size $N$; internal neighborhood size $T$; neighborhood selection probability $\beta$;}
	\KwOut{final sub-population set $\mathcal{P}$\;}
	\tcp{Algorithm \ref{initialization}}
	[$W$, $B$, $\widetilde{B}$, $\Phi$, $\mathcal{P}$, $Z$] := \textit{Initialization}($N$, $T$)\;
	\While{stopping criterion is not met}{
		$\mathcal{U}$ := $\bigcup\limits_{i=1}^{K}\mathcal{P}^{i}$ \tcp*{joint and shuffling}
		\ForEach{$\mathbf{x}$ in $\mathcal{U}$}{
			$cur \leftarrow$ task index where $\mathbf{x}$ is located\;
			$\tau \leftarrow$ subproblem index of individual $\mathbf{x}$\;
			\tcp{Algorithm \ref{neighborhood selection}}
			[$tar$, $Q$] := \textit{CandidateSetSelection}($cur$, $\tau$, $\beta$)\;
			\tcp{Algorithm \ref{reproduction}}
			$\mathbf{\widehat{x}}$ := \textit{Reproduction}($\mathbf{x}$, $tar$, $Q$, $\mathcal{P}$)\;
			\tcp{Algorithm \ref{update}}
			[$Z$, $\mathcal{P}$, $\Phi$, $\widetilde{B}$] := \textit{Update}($cur$, $tar$, $\mathbf{\widehat{x}}$, $\tau$, $Z$, $W$, $Q$, $\mathcal{P}$, $\Phi$, $\widetilde{B}$)\;
		}
	}
	\Return{$\mathcal{P}$}\;
\end{algorithm}

\subsubsection{Initialization} \label{s311}

We use superscripts to symbolize the index of tasks that an item is associated with.
For task $k$, the items to be initialized include: the set of weight vectors $W^k$, the internal neighborhood structure $B^k$ of each subproblem, the task index $\Phi^k$ of the external neighborhood of each subproblem, the external neighborhood structure $\widetilde{B}^k$ of each subproblem, and the sub-population $\mathcal{P}^k$ and the ideal point $Z^k$.
For the weight vectors, we use the method mentioned in \cite{Blank2020} to initialize them. The internal neighborhood of each subproblem is defined as the $T$ closest weight vectors based on the Euclidean distance between the weight vectors \cite{Zhang2007}.
And its external neighborhood, at the initialization step, is defined as all the subproblems of a randomly selected task.
All individuals are encoded within the \emph{unified search space} \cite{Gupta2016}.
Note that when an individual is evaluated by a task, e.g., task $k$, it is necessary to map the individual from the \emph{unified search space} to its decision space, as shown in formula (\ref{map}):
\begin{equation}
	\label{map}
	y_j^k = L_j^k + (U_j^k - L_j^k)*x_j^k
\end{equation}
where $L_j^k$ and $U_j^k$ represent the lower and upper bounds of the $j$-th decision variable of task $k$, respectively, $j \in \{1, 2, \ldots, D_k\}$, and $x_j^k$ denotes the value of the $j$-th dimension of the individual in the \emph{unified search space}.
Finally, the ideal point $Z^k$ is initialized. The main procedure of the initialization process is presented in Algorithm \ref{initialization}.

\begin{algorithm}
	\caption{\textit{Initialization}}
	\label{initialization}
	\KwIn{sub-population size $N$, internal neighborhood size $T$\;}
	\KwOut{weight vectors $W$, internal neighborhood $B$, external neighborhood $\widetilde{B}$, task index of external neighborhood $\Phi$, population set $\mathcal{P}$, ideal point $Z$\;}
	\For{k := 1..$K$}{
		$W^k$ := $\{\mathbf{w}_1^{k}, \mathbf{w}_2^{k}, \ldots, \mathbf{w}_N^{k}\} \leftarrow$ Generate a set of weight vectors by the method proposed in \cite{Blank2020}\;
		$B^k$ := $\{B_1^{k}, B_2^{k}, \ldots, B_N^{k}\}$, where $B_i^{k}$ represents the index set of the $T$ weight vectors that are closest to $\mathbf{w}_i^{k}$ in $W^k$\;
		$\Phi^k$ := $\{\phi_1^{k}, \phi_2^{k}, \dots, \phi_N^{k}\}$, where $\phi_{i}^{k}$ is randomly select from set \{1, 2, \ldots, $K$\}, and $\phi_{i}^{k} \neq k$ \;
		$\widetilde{B}^k$ := $\{\widetilde{B}_1^{k}, \widetilde{B}_2^{k}, \ldots, \widetilde{B}_N^{k}\} $, where $\widetilde{B}_i^{k}$ := \{1, 2, \ldots, $N$\}\;
		$\mathcal{P}^k$ := $\{\mathbf{x}_1^{k}, \mathbf{x}_2^{k}, \ldots, \mathbf{x}_N^{k}\} \leftarrow$ Randomly generate $N$ individuals to form a sub-population for task $k$\;
		$Z^{k}$ := ($z_1^{k}$, $z_2^{k}$, \dots , $z_{m_k}^{k}$) $\leftarrow$ Initialize the ideal point of task $k$ that is represented by $\mathcal{P}^k$\;
	}
	\Return{[$W$, $B$, $\widetilde{B}$, $\Phi$, $\mathcal{P}$, $Z$]}\;
\end{algorithm}

\subsubsection{Candidate Set Selection} \label{s312}

\begin{algorithm}
	\caption{\textit{CandidateSetSelection}}
	\label{neighborhood selection}
	\KwIn{current task index $cur$, current subproblem index $\tau$, neighborhood selection probability $\beta$;}
	\KwOut{target task index $tar$, candidate set $Q$\;}
	$tar$ := $cur$\;
	$Q$ := \{1, 2, \ldots, $N$\}\;
	\If{$rand(0,1) < \beta$}{
		\tcp{randomly select a neighborhood}
		\If{$rand(0,1) < 0.5$}{
			$Q$ := $B_{\tau}^{cur}$ \tcp*{internal neighborhood}
		} \Else{
			$tar$ := $\phi_{\tau}^{cur}$\;
			$Q$ := $\widetilde{B}_{\tau}^{cur}$ \tcp*{external neighborhood}
		}
	}
	\Return{[$tar$, $Q$]}\;
\end{algorithm}

\begin{algorithm}
	\caption{\textit{Reproduction}}
	\label{reproduction}
	\KwIn{individual $\mathbf{x}$, index of target task $tar$, candidate set $Q$, population set $\mathcal{P}$\;}
	\KwOut{offspring $\widehat{\mathbf{x}}$\;}
	$\mathbf{x}_1$ := $\mathbf{x}$\;
	$\mathbf{u} \leftarrow$ Randomly select two individuals from $\mathcal{P}^{tar}(Q)$, which are denoted as $\mathbf{x}_2$ and $\mathbf{x}_3$, generate an offspring $\mathbf{u}$ through the differential evolution operator (DE-rand/1/bin) based on $\mathbf{x}_1$, $\mathbf{x}_2$, and $\mathbf{x}_3$\;
	$\widehat{\mathbf{x}} \leftarrow$ Polynomial mutation operator is employed on $\mathbf{u}$ to generate the offspring \;
	\Return{$\widehat{\mathbf{x}}$}\;
\end{algorithm}

Here, the neighborhood selection probability $\beta$ is used to control whether the candidate set comes from the sub-population of the current task or the neighborhood of the current subproblem. 
If it comes from the neighborhood, then it is randomly selected from the internal neighborhood and external neighborhood of the current subproblem, as shown in lines 3-8 of Algorithm \ref{neighborhood selection}.

\subsubsection{Reproduction} \label{s313}

We employ the differential evolution (DE) \cite{Storn1997} crossover operator to generate the offspring, as shown in Algorithm \ref{reproduction}.
Here, $\mathcal{P}^{tar}(Q)$ denotes the set of individuals whose index values belong to $Q$ in the sub-population $\mathcal{P}^{tar}$ which are associated with task $tar$.
Finally, the offspring is obtained by the polynomial mutation (PM) \cite{Deb2008} operator.

\subsubsection{Update} \label{s314}

\begin{algorithm}
	\caption{\textit{Update}}
	\label{update}
	\KwIn{current task index $cur$, target task index $tar$, offspring $\widehat{\mathbf{x}}$, subproblem index $\tau$, ideal point $Z$, weight vector $W$, candidate set $Q$, population set $\mathcal{P}$, task index of external neighborhood $\Phi$, external neighborhood $\widetilde{B}$\;}
	\KwOut{updated ideal point $Z$, population $\mathcal{P}$, task index of external neighborhood $\Phi$, and external neighborhood $\widetilde{B}$\;}
	\tcp{step 1: update candidate set}
	\If{$cur \neq tar$}{
		\tcp{randomly select a neighborhood}
		\If{$rand(0,1) < 0.5$}{
			$Q$ := $B_{\tau}^{cur}$\;
			$tar$ := $cur$\;
		}
	}
	Evaluate $\widehat{\mathbf{x}}$ by task $tar$\;
	Update ideal point $Z^{tar}$\;
	$A$ := $\emptyset$ \tcp*{record the updated subproblem}
	\ForEach{$q$ in $Q$}{
		\If{$u$($\mathbf{F}(\widehat{\mathbf{x}})$, $\mathbf{w}_{q}^{tar}$) $<$ $u$($\mathbf{F}(\mathbf{x}_{q}^{tar})$, $\mathbf{w}_{q}^{tar}$)}{
			$\mathbf{x}_q^{tar}$ := $\widehat{\mathbf{x}}$\;
			$A$ := $A \cup \{q\}$\;
		}
	}
	\tcp{step 2: update external neighborhood}
	\If{$cur \neq tar$}{
		\If{$|A|$ = 0}{
			$\phi_{\tau}^{cur} \leftarrow$ Randomly select an element from \{1, 2, \ldots, $K$\}, and $\phi_{\tau}^{cur} \neq cur$\;
			$\widetilde{B}_{\tau}^{cur}$ := \{1, 2, \ldots, $N$\}\;
		} \Else{
			$Y \leftarrow$ Set the mean value of the decision variables of $\mathcal{P}^{cur}$($B_{\tau}^{cur}$) as the reference sequence\;
			$X$ := \{$\bar{\mathbf{x}}_{i}^{tar} \mid i \in A$\}, set the compared sequences, where $\bar{\mathbf{x}}_{i}^{tar}$ is the mean value of the decision variables of $\mathcal{P}^{tar}$($B_{A_i}^{tar}$)\;
			$n$ := $\mathop{\arg\max}\limits_{i=1}^{|X|}$ $r_i$, where $r_i$ is the grey relational degree calculated by (\ref{gra}) \;
			$\widetilde{B}_{\tau}^{cur}$ := $B_{A_n}^{tar}$\;
		}
	}
	\Return{[$Z$, $\mathcal{P}$, $\Phi$, $\widetilde{B}$]}\;
\end{algorithm}

Specifically, the update process is organized into two parts. One is to update the candidate set of the target task, as shown in lines 1-11 of Algorithm \ref{update}, and the other is to update the external neighborhood of the current subproblem, as shown in lines 12-20 of Algorithm \ref{update}.
Here, if the external neighborhood of a subproblem participates in the production of the offspring, the updated candidate set is chosen at random from the internal and external neighborhoods of that subproblem, as in lines 1-4 of Algorithm \ref{update}. It can be called a bidirectional update.
Next, the target task is used to evaluate the offspring and the updating of the ideal point of the target task, as shown in lines 5-6.
Lines 8-11 are the specific steps for updating the candidate set, and Achievement Scalarizing Function (ASF) \cite{Pescador2017} is used as the scalarizing function, and its expression is shown in (\ref{asf}):
\begin{eqnarray}
	u^{ASF}(\mathbf{F(x)};\mathbf{w})= \max\limits_{i=1}^{m} ~~ \frac{1}{w_i} |f_i(\mathbf{x}) - z^*_i|
	\label{asf}
\end{eqnarray}

The index set $A$ is used to record which subproblems have been updated by the newly generated offspring.
When the current task and the target task are inconsistent, those subproblems that can bring benefits are worth further mining the valuable information between them. Here, that is, the subproblem in set $A$ and the current subproblem are more worth exploring, and the corresponding tasks have more investment value.
If $A$ is an empty set, i.e., the current subproblem fails to update any of the subproblems within its external neighborhood, then the external neighborhood of the current subproblem is reset, as shown in lines 13-15 of Algorithm \ref{update}.
Otherwise, we use the grey relation analysis to select the internal neighborhood of a subproblem (in set $A$) in the target task with the largest correlation value as the external neighborhood of the current subproblem, as shown in lines 17-20 of Algorithm \ref{update}.

\subsection{Discussion} \label{s34} 

In our proposed algorithm, the internal neighborhood constructs the relationship between subproblems within the same optimization task. Whereas, the external neighborhood constructs the relationship between the subproblems belonging to different tasks.
For a subproblem, its internal neighborhood is determined by the pre-defined weight vectors, and the weight vectors fix the internal neighborhood structure.
Whereas its external neighborhood is explored and mined through the exchange of information between tasks, and will change dynamically with the evolutionary process of the population.
When the optimization of a subproblem of the current task can give a promotion to a subproblem of the target task, then the neighborhood of this subproblem in the target task is also more worthy to be explored and mined.
Of course, for a stochastic algorithm, it is difficult to guarantee that each exchange of information between tasks will get a significant reward. 
That is, individuals in the candidate set (external neighborhood) might not be updated by their offspring.
In this case, then, taking all subproblems of a task as the external neighborhood of the current subproblem can make the communication between tasks more exploratory.
Besides, we consider the internal and external neighborhoods of the subproblem to be equally important, and this is the reason that the candidate set to be updated is randomly selected (lines 2-4 of Algorithm \ref{update}) from the internal and external neighborhoods of the subproblem when there is an exchange of information between different tasks. A more detailed discussion of internal neighborhood and external neighborhood is presented in Section \ref{s46}.

\section{Experimental Studies} \label{s4}

In this section, we first compare our proposed algorithm with five state-of-the-art algorithms. Then, the effect of internal and external neighborhoods on our proposed algorithm is further analyzed. Finally, sensitivity analysis experiments are conducted on some parameters of our proposed algorithm.

\subsection{Competing Algorithms} \label{s41}

Four state-of-the-art algorithms, namely MO-MFEA \cite{Gupta2017}, MO-MFEA-II \cite{Bali2020_2}, EMTIL \cite{Lin2019}, MFEA/D-DE \cite{Yao2020} and a traditional decomposition-based multiobjective evolutionary algorithm, MOEA/D \cite{Li2009}, are used as the compared algorithms.
Among them, MO-MFEA is the first paradigm of the multiobjective multitasking evolutionary algorithm.
MO-MFEA-II is a variant of MO-MFEA which employs data-driven technology to establish a similar relationship model between different tasks. This model is used to adjust the frequency of information transfer in order to guarantee that useful information can be fully utilized and at the same time useless information can be abandoned.
Incremental learning method is utilized in EMTIL to exploit potentially valuable information between different tasks.
The selection pressure in the above three algorithms is based on the dominant strategy.
MFEA/D-DE is the first attempt to use the decomposition-based strategy in MOMTO.
Our algorithm is also based on a decomposition strategy, so here, the traditional decomposition-based multiobjective evolutionary algorithm MOEA/D \cite{Li2009} is also used as a compared algorithm.

\subsection{Test Instances} \label{s42}

We evaluate the MTEA/D-DN algorithm on 9 multiobjective MTO benchmark test instances, and each test instance is composed of two MOPs.
In terms of similarity between the tasks of a test instance, these test instances can be categorized into three groups: high similarity (HS), medium similarity (MS) and low similarity (LS).
From the intersection of the global minima, they can be categorized into complete intersection (CI), partial intersection (PI) and no intersection (NI).
Combining these two perspectives together, these 9 test instances can be denoted as CIHS, CIMS, CILS, PIHS, PIMS, PILS, NIHS, NIMS and NILS. Taking CIHS as an example, it indicates that the test instance is Complete Intersection with High Similarity.
For the detailed settings of these test instances, such as the variable range and dimension of each task, etc., please refer to the literature \cite{Yuan2017}.

\subsection{Performance Metrics} \label{s43}

The most straightforward manner of evaluating the performance of a multitasking optimization algorithm is to analyze the quality of solutions of each task independently. For an MOP, it is difficult to comprehensively measure the performance of an algorithm with a single metric.
In general, the diversity and convergence of the approximate PF obtained by the algorithm are two important metrics to measure the performance of an algorithm. The inverse generation distance (IGD) \cite{schuetze2010} and Hypervolume (HV) \cite{Emmerich2005} are both composite metrics that measure the quality of the approximate PF obtained by the algorithm.
Herein, the formula for calculating the IGD metric can be described in the following form: 
\begin{eqnarray}
	\begin{aligned}
		{\rm IGD}(S, PF^{*}) = \frac{1}{|PF^{*}|} \sqrt{\sum\limits_{\mathbf{x} \in PF^{*}} (\min\limits_{\mathbf{y} \in S} dist(\mathbf{x}, \mathbf{y}))^{2}} \\
	\end{aligned}
	\label{IGD metric}
\end{eqnarray}
where, $S$ represents the set of approximate PF obtained by the algorithm, and $PF^*$ represents the subset of true PF. $dist(\mathbf{x}, \mathbf{y})$ is the Euclidean distance between points $\mathbf{x}$ and $\mathbf{y}$.
A smaller IGD metric means a better performance of the algorithm.

Then, given a reference point $\mathbf{z}^r = (z_1^r, z_2^r, \ldots, z_m^r)$, the Hypervolume metric can be computed by using formula (\ref{HV metric}).
\begin{eqnarray}
	\begin{aligned}
		{\rm HV}(S, \mathbf{z}^{r}) = {\rm VOL} \left(\bigcup_{\mathbf{y} \in S}[\mathbf{y}_1,z_1^{r}]\times \ldots \times[\mathbf{y}_m,z_m^{r}] \right)
		\label{HV metric}
	\end{aligned}
\end{eqnarray}
where, VOL($\cdot$) represents a Lebesgue measure, and nadir point can be used as reference point or defined by the user.
The HV metric is the opposite of the IGD metric, where a larger value indicates a better performance of the algorithm.

Note that for both metrics, we have normalized the points on the approximate PF obtained by the algorithm by using the nadir point and ideal point derived from the true PF. For each test instance, all algorithms are run independently 21 times.

\subsection{Parameter Settings} \label{s44}

For the same test instance, the termination condition for all algorithms is the number of evaluations ($Eva$) of the objective function, which is set to $K \times 50,000$, where $K$ is the number of tasks. Here, the number of evaluations of the objective function refers to the sum of the number of evaluations of all tasks in a test instance.

\begin{table}[!htbp]
	\caption{Operator Parameters}
	\label{parameters1}
	\centering
	\renewcommand\arraystretch{1.1}
	\begin{tabular}{lllllll}
	\hline & \multicolumn{2}{c}{SBX} & \multicolumn{2}{c}{DE} & \multicolumn{2}{c}{PM} \\
	\cline{2-3} \cline{4-5} \cline{6-7} & $\eta_c$ & $p_c$ & $F$ & $Cr$ & $\eta_m$ & $p_m$ \\
	\hline 
	MO-MFEA & $10$ & $1.0$ & -- & -- & $10$ & $1/D$ \\
	MO-MFEA-II & $10$ & $1.0$ & -- & -- & $10$ & $1/D$ \\
	EMTIL & $15$ & $1.0$ & -- & -- & $20$ & $1/D$ \\
	MFEA/D-DE & -- & -- & $0.5$ & $0.9$ & $20$ & $1/D$ \\
	MOEA/D & -- & -- & $0.5$ & $0.9$ & $20$ & $1/D$ \\
	MTEA/D-DN & -- & -- & $0.5$ & $0.9$ & $20$ & $1/D$ \\
	\hline
	\end{tabular}
\end{table}

For the reproduction of offspring, the MO-MFEA, MO-MFEA-II and EMTIL algorithms apply the simulated binary crossover (SBX) operator \cite{Agrawal1995}, and the MFEA/D-DE, MOEA/D and MTEA/D-DN algorithms apply the DE \cite{Storn1997} crossover operator. Moreover, the final offspring in all algorithms are obtained from PM \cite{Deb2008} operator. The parameter settings of these operators for producing offspring are shown in Table \ref{parameters1}. Here, $\eta_c$ and $p_c$ are the distribution index and crossover probability for SBX operator, $F$ and $Cr$ are constant factor and crossover constant for DE operator, and $\eta_m$ and $p_m$ are the distribution index and mutation probability for PM operator.
The symbol $D$ represents the dimension of the decision variable in the \emph{unified search space} (except for MOEA/D, which is the dimension of the decision variable of the optimized task).

\begin{table}[!htbp]
		\caption{Decomposition Strategy parameters}
		\label{parameters2}
		\centering
		\renewcommand\arraystretch{1.1}
		\begin{tabular}{llll}
		\hline & $\beta$ & $T$ & $n_r$ \\
		\hline 
		MFEA/D-DE & $0.8$ & $10$ & $2$ \\
		MOEA/D & $0.9$ & $20$ & $2$ \\
		MTEA/D-DN & $0.1$ & $10$ & -- \\
		\hline
		\end{tabular}
\end{table}

In addition, the special parameters of some algorithms, such as the random mating probability $rmp$ in the MO-MFEA and MFEA/D-DE algorithms, are set to 0.3 and 0.1 respectively. The number of transferred solutions in EMTIL is set to 10. The settings of neighborhood selection probability $\beta$, neighborhood size $T$, and the maximum number of replacement $n_r$ in MFEA/D-DE, MOEA/D and our proposed algorithm are shown in Table \ref{parameters2}.

\subsection{Compared to State-of-the-Art Algorithms} \label{s45}

Tables \ref{IGD state-of-the-art} and \ref{HV state-of-the-art} show the results of the statistical analysis of the means and standard deviations of our proposed algorithm and its rival algorithms with respect to the IGD and HV metrics, respectively.
Here, the algorithm that achieves the best performance is marked with dark gray shading and the one that achieves the second best is marked with light gray shading.
On the whole, it is clear that, for both IGD and HV metrics, the statistical results demonstrate that our proposed algorithm performs better than its rival algorithms.
Since all the algorithms failed to obtain a solution that dominates the reference point each time they were run, the HV metrics corresponding to PIHS2, PIMS2, NIHS1, and NILS2 are all 0 in Table \ref{HV state-of-the-art}.
The multitasking evolutionary algorithm MFEA/D-DE, also based on the decomposition strategy, performs second best, with the light gray shading covering the most.
This means that the decomposition-based multiobjective evolutionary algorithms are more preferred to solve the MOMTO problems.

\begin{figure*}
	\centering
	\includegraphics[width=6.3 in]{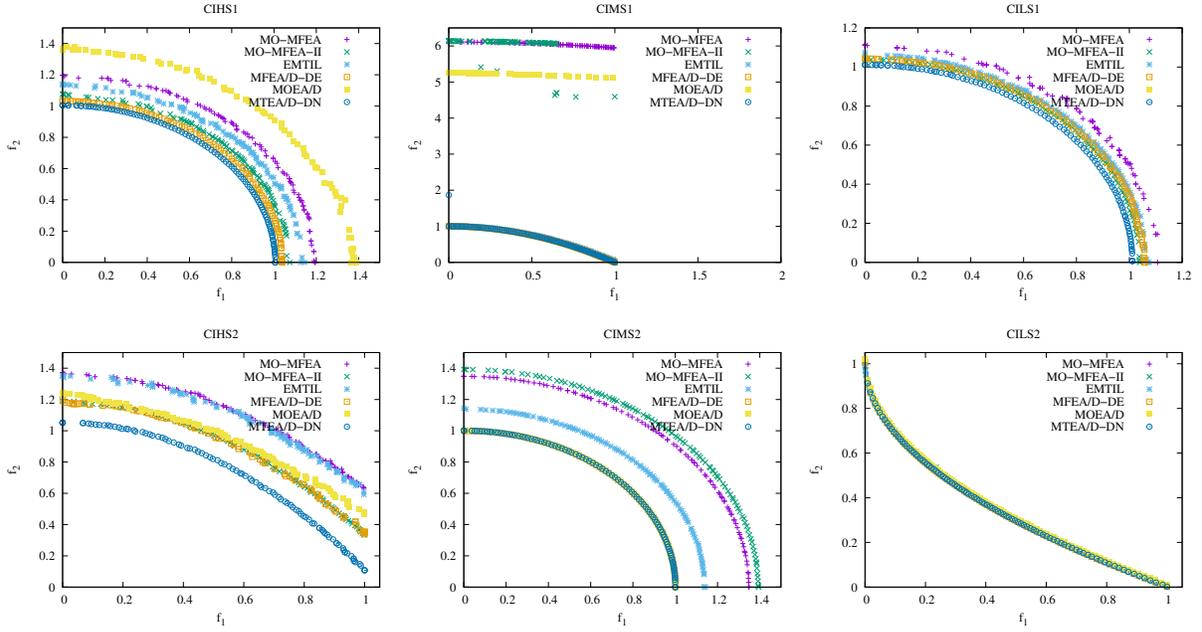}
	\caption{Non-dominated solution set of median IGD metric obtained by MTEA/D-ND and its rival algorithms for CIHS, CIMS and CILS}
	\label{CIHS-CIMS-CILS-IGD}
\end{figure*}

Taking the test instances of CIHS, CIMS and CILS as examples, we present the final non-dominated solution sets of all testing algorithms in terms of the median IGD metric in Figure \ref{CIHS-CIMS-CILS-IGD}. From this figure, it can be found that our proposed MTEA/D-DN can always achieve the best or competitive results.

\begin{table*}[!htbp]
	\caption{The Performance Analysis of MTEA/D-DN and Its 5 State-of-the-Art Comparison Algorithms in Terms of Mean and Standard Deviation of IGD Metric}
	\label{IGD state-of-the-art}
	\centering
	\renewcommand\arraystretch{1.3}
	\begin{tabular}{ccccccc}
	\hline & MO-MFEA & MO-MFEA-II & EMTIL & MFEA/D-DE & MOEA/D &  MTEA/D-DN\\
	\hline 
	CIHS1 & $  5.82e-03_{ 1.6e-03}$ & $  1.99e-03_{ 8.2e-04}$ & $  4.69e-03_{ 1.7e-03}$ & \cellcolor{graylt}$  1.23e-03_{ 2.8e-04}$ & $  1.10e-02_{ 2.4e-03}$ & \cellcolor{graydk}$  2.86e-04_{ 7.4e-05}$ \\
	CIHS2 & $  1.17e-02_{ 2.2e-03}$ & $  5.56e-03_{ 1.4e-03}$ & $  1.06e-02_{ 1.9e-03}$ & \cellcolor{graylt}$  5.55e-03_{ 8.8e-04}$ & $  8.61e-03_{ 6.1e-03}$ & \cellcolor{graydk}$  1.70e-03_{ 2.8e-04}$ \\
	\hline
	CIMS1 & $  1.12e-01_{ 1.0e-01}$ & $  1.18e-01_{ 8.2e-02}$ & $  4.38e-02_{ 7.3e-02}$ & \cellcolor{graylt}$  1.50e-04_{ 3.4e-05}$ & $  1.34e-01_{ 4.9e-02}$ & \cellcolor{graydk}$  1.45e-04_{ 6.2e-06}$ \\
	CIMS2 & $  1.32e-02_{ 1.1e-02}$ & $  1.30e-02_{ 7.9e-03}$ & $  7.87e-03_{ 8.8e-03}$ & $  2.13e-04_{ 1.3e-04}$ & \cellcolor{graylt}$  1.82e-04_{ 1.1e-05}$ & \cellcolor{graydk}$  1.82e-04_{ 1.3e-05}$ \\
	\hline
	CILS1 & $  3.47e-03_{ 6.9e-04}$ & \cellcolor{graylt}$  1.45e-03_{ 3.9e-04}$ & $  2.14e-03_{ 4.5e-04}$ & $  1.62e-03_{ 3.9e-04}$ & $  1.10e+00_{ 6.9e-01}$ & \cellcolor{graydk}$  4.74e-04_{ 9.4e-05}$ \\
	CILS2 & $  2.37e-04_{ 8.1e-06}$ & \cellcolor{graylt}$  1.91e-04_{ 8.9e-06}$ & $  2.27e-04_{ 1.0e-05}$ & \cellcolor{graydk}$  1.83e-04_{ 5.0e-06}$ & $  3.55e-04_{ 2.7e-05}$ & $  2.05e-04_{ 4.0e-06}$ \\
	\hline
	PIHS1 & $  4.41e-02_{ 1.5e-02}$ & $  1.91e-02_{ 8.0e-03}$ & $  2.69e-02_{ 6.7e-03}$ & $  6.04e-03_{ 1.7e-03}$ & \cellcolor{graylt}$  4.97e-03_{ 1.4e-03}$ & \cellcolor{graydk}$  1.06e-03_{ 3.0e-04}$ \\
	PIHS2 & $  1.26e+00_{ 2.1e-01}$ & $  9.00e-01_{ 1.5e-01}$ & \cellcolor{graylt}$  6.82e-01_{ 1.2e-01}$ & $  1.68e+00_{ 4.2e-01}$ & $  2.29e+00_{ 4.4e-01}$ & \cellcolor{graydk}$  1.18e-01_{ 3.7e-02}$ \\
	\hline
	PIMS1 & $  5.13e-03_{ 1.9e-03}$ & $  3.96e-03_{ 1.3e-03}$ & \cellcolor{graylt}$  3.31e-03_{ 6.4e-04}$ & $  7.60e-03_{ 3.0e-03}$ & $  9.34e-03_{ 3.1e-03}$ & \cellcolor{graydk}$  2.08e-03_{ 8.5e-04}$ \\
	PIMS2 & $  1.37e+01_{ 3.4e+00}$ & \cellcolor{graylt}$  1.29e+01_{ 3.8e+00}$ & \cellcolor{graydk}$  7.43e+00_{ 1.9e+00}$ & $  2.09e+01_{ 5.5e+00}$ & $  2.05e+01_{ 7.6e+00}$ & $  1.56e+01_{ 6.8e+00}$ \\
	\hline
	PILS1 & $  1.55e-03_{ 3.9e-04}$ & $  7.19e-04_{ 1.8e-04}$ & $  5.70e-04_{ 1.5e-04}$ & $  6.25e-04_{ 2.2e-04}$ & \cellcolor{graylt}$  5.41e-04_{ 1.8e-04}$ & \cellcolor{graydk}$  3.98e-04_{ 2.3e-04}$ \\
	PILS2 & $  7.76e-02_{ 7.0e-03}$ & \cellcolor{graylt}$  5.76e-02_{ 8.3e-03}$ & \cellcolor{graydk}$  3.79e-02_{ 1.0e-02}$ & $  5.34e-01_{ 1.9e-01}$ & $  4.21e-01_{ 2.6e-01}$ & $  7.96e-02_{ 1.1e-01}$ \\
	\hline
	NIHS1 & $  2.96e+00_{ 6.0e-01}$ & $  2.22e+00_{ 2.8e-01}$ & $  2.97e+00_{ 5.3e-01}$ & \cellcolor{graylt}$  1.79e+00_{ 5.7e-02}$ & $  1.57e+01_{ 2.5e+01}$ & \cellcolor{graydk}$  1.50e+00_{ 2.2e-02}$ \\
	NIHS2 & $  9.71e-03_{ 3.9e-03}$ & $  4.49e-03_{ 1.6e-03}$ & $  8.70e-03_{ 2.9e-03}$ & \cellcolor{graylt}$  1.86e-03_{ 4.0e-04}$ & $  3.15e-03_{ 1.0e-03}$ & \cellcolor{graydk}$  4.51e-04_{ 1.1e-04}$ \\
	\hline
	NIMS1 & $  5.77e-01_{ 3.9e-01}$ & $  4.92e-01_{ 3.6e-01}$ & $  4.60e-01_{ 3.5e-01}$ & \cellcolor{graydk}$  1.50e-01_{ 9.6e-03}$ & $  2.05e-01_{ 1.8e-01}$ & \cellcolor{graylt}$  1.66e-01_{ 2.5e-01}$ \\
	NIMS2 & $  2.18e-01_{ 2.7e-01}$ & $  1.40e-01_{ 2.1e-01}$ & $  1.27e-01_{ 2.3e-01}$ & \cellcolor{graylt}$  8.76e-04_{ 1.3e-03}$ & $  2.88e-02_{ 2.6e-02}$ & \cellcolor{graydk}$  6.96e-04_{ 6.6e-04}$ \\
	\hline
	NILS1 & $  2.85e-03_{ 1.7e-03}$ & $  1.08e-03_{ 2.6e-04}$ & $  9.17e-04_{ 1.1e-04}$ & \cellcolor{graylt}$  7.29e-04_{ 8.1e-05}$ & \cellcolor{graydk}$  7.00e-04_{ 5.9e-05}$ & $  1.27e-03_{ 6.9e-04}$ \\
	NILS2 & $  6.45e-01_{ 4.2e-04}$ & $  6.45e-01_{ 1.3e-03}$ & $  6.44e-01_{ 3.4e-04}$ & \cellcolor{graydk}$  5.97e-01_{ 1.5e-01}$ & \cellcolor{graylt}$  6.13e-01_{ 1.3e-01}$ & $  6.43e-01_{ 5.0e-04}$ \\
	\hline
	\end{tabular}
\end{table*}

\begin{table*}
	\caption{The Performance Analysis of MTEA/D-DN and Its 5 State-of-the-Art Comparison Algorithms in Terms of Mean and Standard Deviation of HV Metric}
	\label{HV state-of-the-art}
	\centering
	\renewcommand\arraystretch{1.3}
	\begin{tabular}{ccccccc}
	\hline & MO-MFEA & MO-MFEA-II & EMTIL & MFEA/D-DE & MOEA/D &  MTEA/D-DN\\
	\hline 
	CIHS1 & $  5.47e-02_{ 3.0e-02}$ & $  1.40e-01_{ 2.5e-02}$ & $  7.52e-02_{ 3.1e-02}$ & \cellcolor{graylt}$  1.63e-01_{ 9.8e-03}$ & $  8.00e-03_{ 1.0e-02}$ & \cellcolor{graydk}$  2.00e-01_{ 3.8e-03}$ \\
	CIHS2 & $  4.87e-02_{ 2.8e-02}$ & \cellcolor{graylt}$  1.55e-01_{ 3.3e-02}$ & $  6.26e-02_{ 2.2e-02}$ & $  1.50e-01_{ 2.1e-02}$ & $  1.13e-01_{ 4.6e-02}$ & \cellcolor{graydk}$  2.61e-01_{ 9.9e-03}$ \\
	\hline
	CIMS1 & $  8.34e-02_{ 1.3e-01}$ & $  5.25e-02_{ 1.0e-01}$ & $  1.97e-01_{ 1.4e-01}$ & \cellcolor{graylt}$  3.27e-01_{ 2.4e-03}$ & $  0.00e+00_{ 0.0e+00}$ & \cellcolor{graydk}$  3.28e-01_{ 4.6e-04}$ \\
	CIMS2 & $  4.25e-02_{ 6.3e-02}$ & $  2.56e-02_{ 3.9e-02}$ & $  7.98e-02_{ 6.1e-02}$ & $  2.06e-01_{ 7.1e-03}$ & \cellcolor{graylt}$  2.08e-01_{ 1.1e-03}$ & \cellcolor{graydk}$  2.08e-01_{ 1.3e-03}$ \\
	\hline
	CILS1 & $  9.83e-02_{ 1.7e-02}$ & \cellcolor{graylt}$  1.56e-01_{ 1.3e-02}$ & $  1.35e-01_{ 1.3e-02}$ & $  1.50e-01_{ 1.2e-02}$ & $  0.00e+00_{ 0.0e+00}$ & \cellcolor{graydk}$  1.91e-01_{ 4.1e-03}$ \\
	CILS2 & $  6.55e-01_{ 5.4e-04}$ & $  6.58e-01_{ 5.3e-04}$ & $  6.55e-01_{ 6.2e-04}$ & \cellcolor{graylt}$  6.58e-01_{ 4.4e-04}$ & $  6.48e-01_{ 1.3e-03}$ & \cellcolor{graydk}$  6.58e-01_{ 3.7e-04}$ \\
	\hline
	PIHS1 & $  8.21e-04_{ 2.6e-03}$ & $  1.07e-01_{ 1.1e-01}$ & $  2.77e-02_{ 4.7e-02}$ & $  4.10e-01_{ 6.6e-02}$ & \cellcolor{graylt}$  4.54e-01_{ 5.4e-02}$ & \cellcolor{graydk}$  6.16e-01_{ 1.3e-02}$ \\
	PIHS2 & $  0.00e+00_{ 0.0e+00}$ & $  0.00e+00_{ 0.0e+00}$ & $  0.00e+00_{ 0.0e+00}$ & $  0.00e+00_{ 0.0e+00}$ & $  0.00e+00_{ 0.0e+00}$ & $  0.00e+00_{ 0.0e+00}$ \\
	\hline
	PIMS1 & $  7.02e-02_{ 3.2e-02}$ & $  9.18e-02_{ 2.8e-02}$ & \cellcolor{graylt}$  1.05e-01_{ 1.6e-02}$ & $  3.88e-02_{ 3.1e-02}$ & $  2.20e-02_{ 2.4e-02}$ & \cellcolor{graydk}$  1.38e-01_{ 2.5e-02}$ \\
	PIMS2 & $  0.00e+00_{ 0.0e+00}$ & $  0.00e+00_{ 0.0e+00}$ & $  0.00e+00_{ 0.0e+00}$ & $  0.00e+00_{ 0.0e+00}$ & $  0.00e+00_{ 0.0e+00}$ & $  0.00e+00_{ 0.0e+00}$ \\
	\hline
	PILS1 & $  1.52e-01_{ 1.2e-02}$ & $  1.82e-01_{ 6.9e-03}$ & $  1.88e-01_{ 6.2e-03}$ & $  1.85e-01_{ 8.6e-03}$ & \cellcolor{graylt}$  1.89e-01_{ 7.2e-03}$ & \cellcolor{graydk}$  1.96e-01_{ 1.0e-02}$ \\
	PILS2 & $  0.00e+00_{ 0.0e+00}$ & $  0.00e+00_{ 0.0e+00}$ & $  0.00e+00_{ 0.0e+00}$ & $  0.00e+00_{ 0.0e+00}$ & $  0.00e+00_{ 0.0e+00}$ & \cellcolor{graydk}$  2.05e-02_{ 2.9e-02}$ \\
	\hline
	NIHS1 & $  0.00e+00_{ 0.0e+00}$ & $  0.00e+00_{ 0.0e+00}$ & $  0.00e+00_{ 0.0e+00}$ & $  0.00e+00_{ 0.0e+00}$ & $  0.00e+00_{ 0.0e+00}$ & $  0.00e+00_{ 0.0e+00}$ \\
	NIHS2 & $  2.85e-01_{ 1.1e-01}$ & $  4.69e-01_{ 6.6e-02}$ & $  3.14e-01_{ 9.7e-02}$ & \cellcolor{graylt}$  5.80e-01_{ 1.8e-02}$ & $  5.26e-01_{ 4.2e-02}$ & \cellcolor{graydk}$  6.44e-01_{ 5.5e-03}$ \\
	\hline
	NIMS1 & $  0.00e+00_{ 0.0e+00}$ & $  0.00e+00_{ 0.0e+00}$ & $  0.00e+00_{ 0.0e+00}$ & $  0.00e+00_{ 0.0e+00}$ & $  0.00e+00_{ 0.0e+00}$ & \cellcolor{graydk}$  4.20e-05_{ 1.6e-04}$ \\
	NIMS2 & $  3.70e-02_{ 6.6e-02}$ & $  3.85e-02_{ 7.5e-02}$ & $  4.79e-02_{ 8.9e-02}$ & \cellcolor{graylt}$  2.97e-01_{ 4.4e-02}$ & $  6.22e-02_{ 9.2e-02}$ & \cellcolor{graydk}$  3.01e-01_{ 2.7e-02}$ \\
	\hline
	NILS1 & $  1.16e-01_{ 7.7e-02}$ & $  2.85e-01_{ 4.5e-02}$ & $  3.16e-01_{ 2.5e-02}$ & \cellcolor{graylt}$  3.58e-01_{ 2.4e-02}$ & \cellcolor{graydk}$  3.67e-01_{ 1.8e-02}$ & $  2.59e-01_{ 8.3e-02}$ \\
	NILS2 & $  0.00e+00_{ 0.0e+00}$ & $  0.00e+00_{ 0.0e+00}$ & $  0.00e+00_{ 0.0e+00}$ & $  0.00e+00_{ 0.0e+00}$ & $  0.00e+00_{ 0.0e+00}$ & $  0.00e+00_{ 0.0e+00}$ \\
	\hline
	\end{tabular}
\end{table*}

\begin{figure*}
	\centering
	\includegraphics[width=6.3 in]{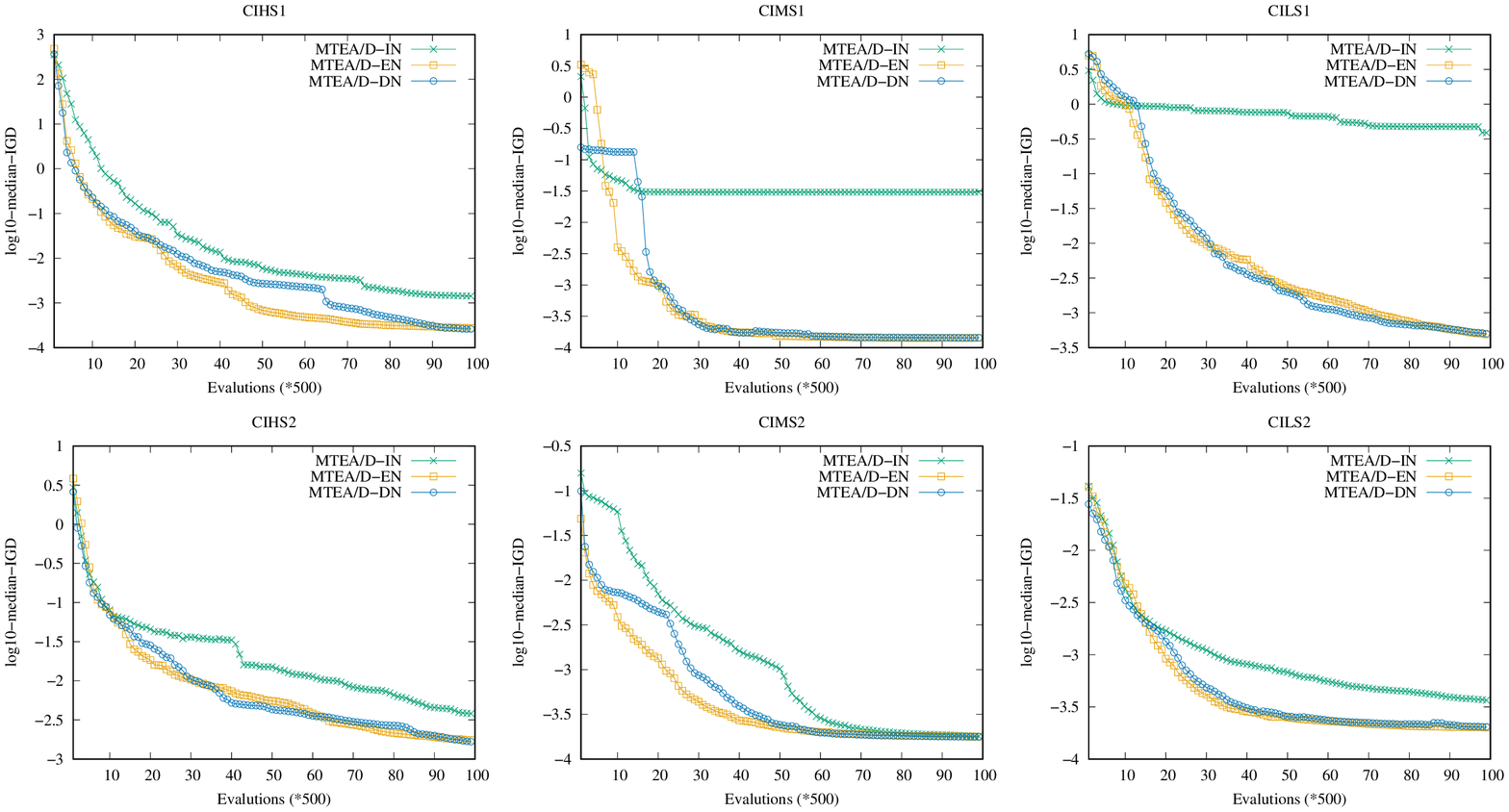}
	\caption{The convergence curves of the median process IGD metric for CIHS, CIMS and CILS}
	\label{CIHS-CIMS-CILS}
\end{figure*}

\subsection{Discussion of the Internal and External Neighborhoods} \label{s46}

To analyze the effect of the proposed dual neighborhood strategy, in this experiment we further discuss the performance of our proposed algorithm under the condition of employing only internal neighborhood, or only external neighborhood.
In the case of employing only the internal neighborhood, we denote the corresponding algorithm as MTEA/D-IN, which means that the information exchange channel between different tasks is closed, and that the external neighborhood of the subproblem is not participated in the candidate set selection and update.
On the other hand, in the case of employing only the external neighborhood, the resulting algorithm is denoted as MTEA/D-EN, which means that the generated offspring will only be used to update its external neighborhood. In this situation there is some information exchange between tasks, but it is unidirectional.
Here, the above 9 test instances are used as benchmark test problems and the IGD metric is adopted as a performance metric. Our proposed algorithm is taken as the compared algorithm and statistical analysis is performed utilizing the Friedman's test. The results of average ranking obtained by the three algorithms are shown in Table \ref{external internal}.
Based on the results, it can be seen that the proposed algorithm performs much better when it has the dual neighborhood, and that the algorithm with only the external neighborhood (i.e., MTEA/D-EN) can achieve a slightly better average ranking than the algorithm with only the internal neighborhood (i.e., MTEA/D-IN).

\begin{table}[!htp]
	\centering
	\renewcommand\arraystretch{1.3}
	\caption{Average Ranking of the Algorithms}
	\begin{tabular}{cc}
	\hline
	Algorithm & Average Ranking\\
	\hline
	MTEA/D-EN & 2.17\\
	MTEA/D-IN & 2.33\\
	MTEA/D-DN & 1.50\\
	\hline
	\end{tabular}
	\label{external internal}
\end{table}

\begin{figure*}
	\centering
	\includegraphics[width=6.3 in]{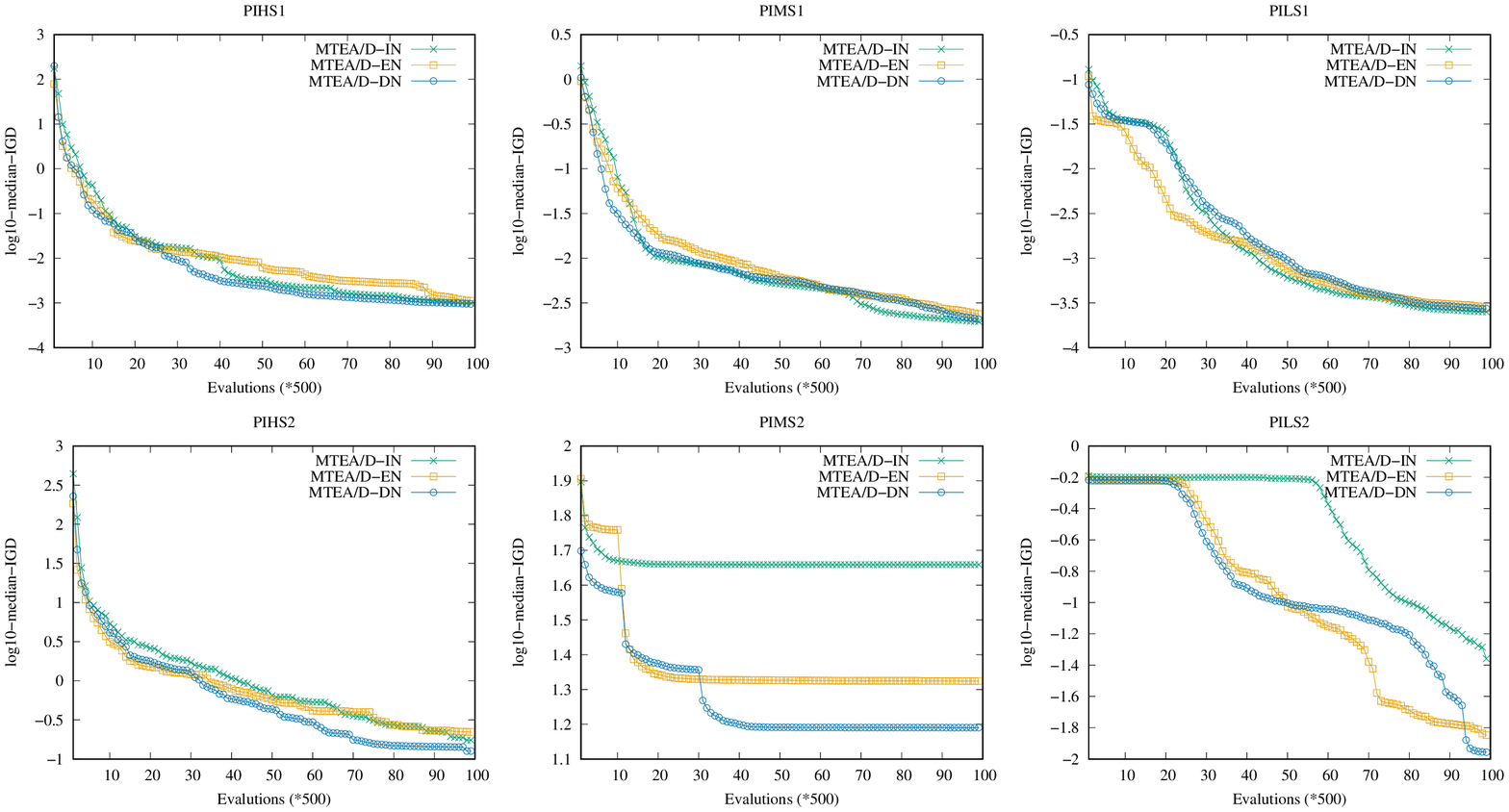}
	\caption{The convergence curves of the median process IGD metric for PIHS, PIMS and PILS}
	\label{PIHS-PIMS-PILS}
\end{figure*}

\begin{figure*}
	\centering
	\includegraphics[width=6.3 in]{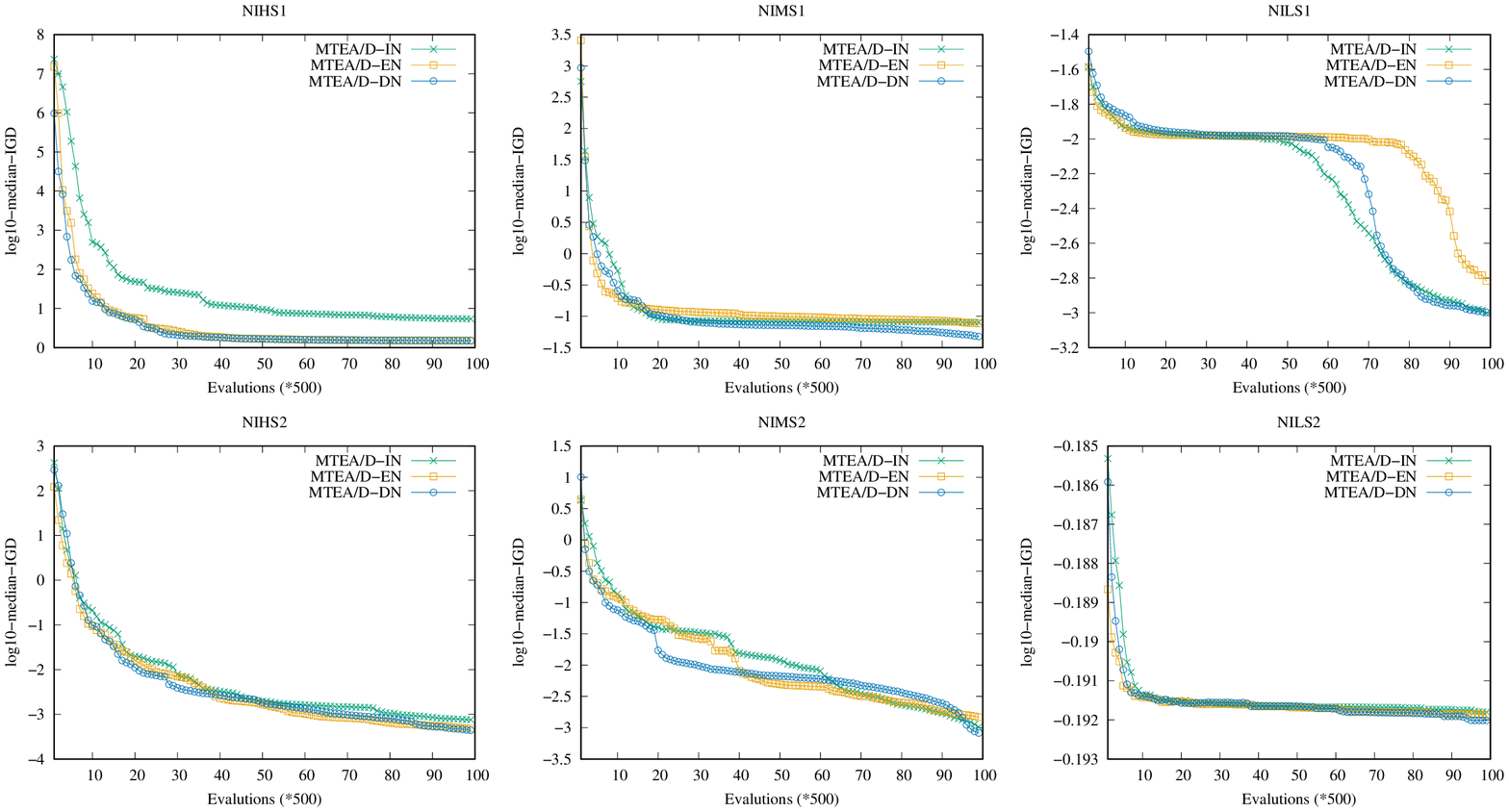}
	\caption{The convergence curves of the median process IGD metric for NIHS, NIMS and NILS}
	\label{NIHS-NIMS-NILS}
\end{figure*}

Further, the convergence curves of the median process IGD metrics for these test instances are shown in Figures \ref{CIHS-CIMS-CILS}-\ref{NIHS-NIMS-NILS}.
From Figure \ref{CIHS-CIMS-CILS} focusing on the tasks with complete intersections, it indicates that the adoption of external neighborhood can significantly improve the convergence speed. The main reason is that the information exchange will become more efficient between subproblems with external neighborhoods because the tasks have complete intersections. From Figure \ref{PIHS-PIMS-PILS} focusing on the tasks with partial intersections, the power of external neighborhood tends to deteriorate, but it is still very efficient for some tasks. With respect to Figure \ref{NIHS-NIMS-NILS} focusing on the tasks with no intersections, the performance difference between the internal neighborhood and the external neighborhood becomes insignificant.
Based on these results, it can be concluded that mining valuable information between tasks (i.e., MTEA/D-DN and MTEA/D-EN) can improve the convergence speed when solving different tasks, especially when the tasks have high intersections between each other.
However, when the internal and external neighborhoods of the subproblem work together, they can make the algorithm perform better and more robust.

\subsection{Sensitivity Analysis} \label{s47}

\begin{figure}
	\centering
	\subfigure[]{
		\includegraphics[width=3 in]{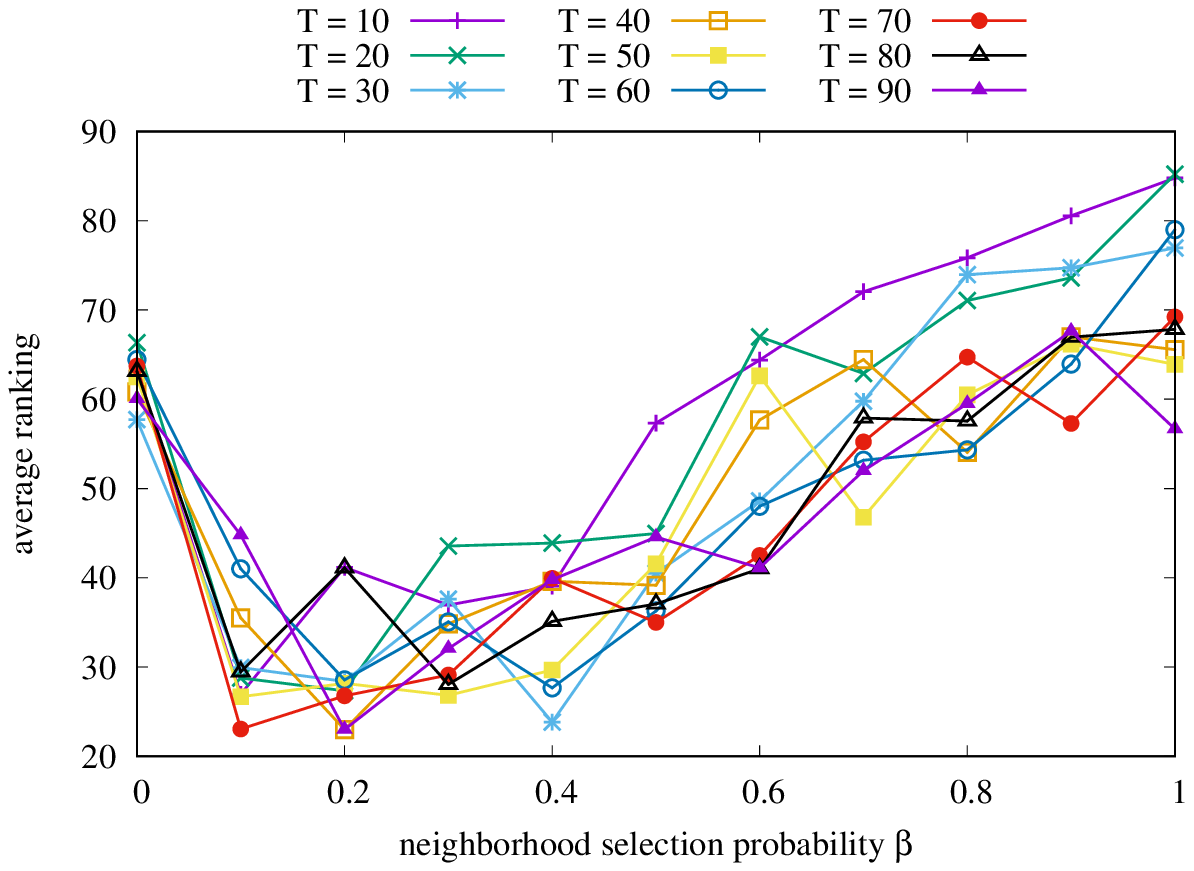}
		\label{beta1-n-FriedmanTestIGD2}}
	\subfigure[]{
		\includegraphics[width=3 in]{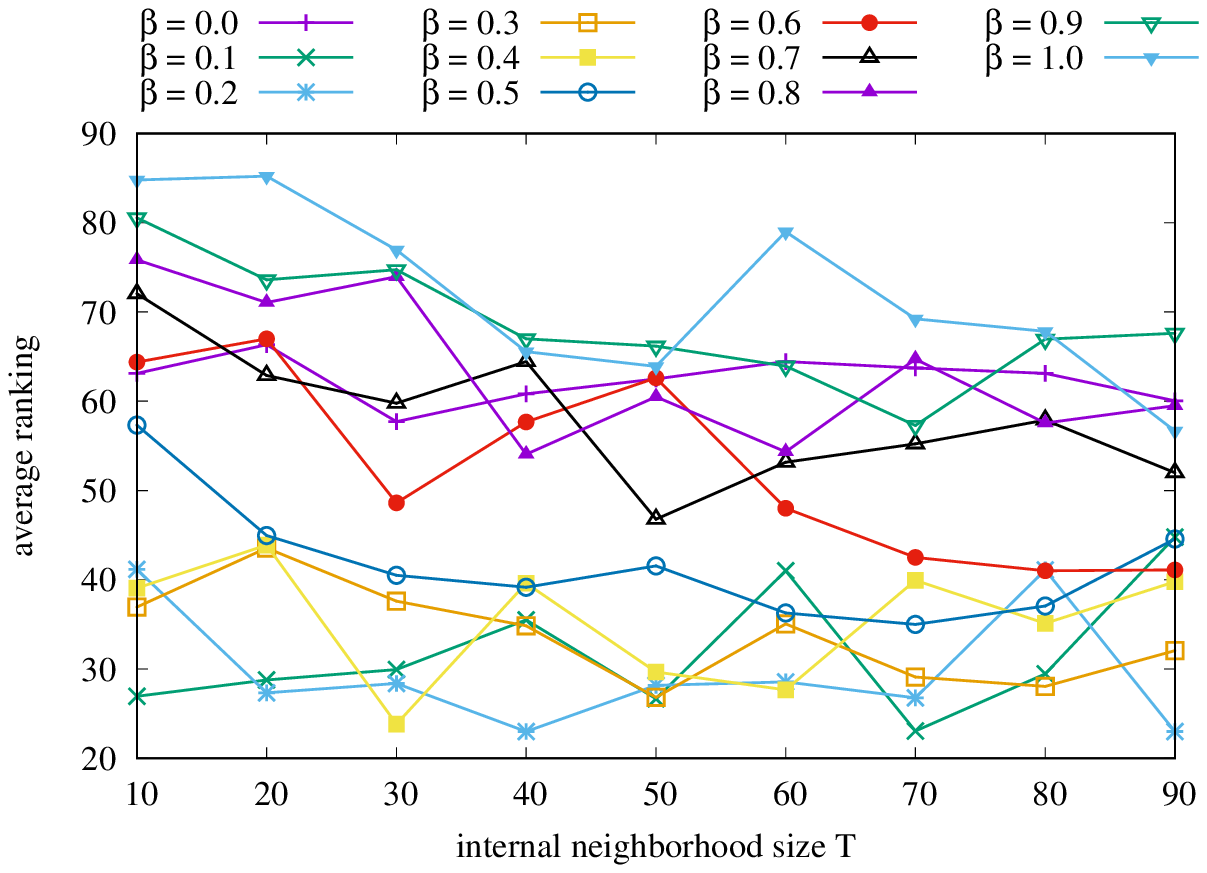}
		\label{beta1-n-FriedmanTestIGD}}
	\caption{Average ranking with different hyperparameters under the IGD metric}
	\label{sensitivity}
\end{figure}

In this experiment, we analyze the sensitivity of two hyperparameters in our proposed algorithm, i.e., the probability of neighborhood selection $\beta$ and the internal neighborhood size $T$, on the performance of the proposed algorithm.
The values of the two parameters are set as $\beta \in \{0.0, 0.1, \ldots, 1.0\}$, and $T \in \{10, 20, \ldots, 90\}$, respectively, and the other parameters are the same as the above settings. Based on the test instances mentioned before, the Friedman's test is adopted for statistical analysis.
Figure \ref{sensitivity} illustrates the change of average ranking obtained by different values of $\beta$ and $T$.
Here, Figure \ref{beta1-n-FriedmanTestIGD2} shows that for a certain internal neighborhood size, the performance of the algorithm (in terms of average ranking) first gets better but then worse as the increase of the neighborhood selection probability.
This indicates that when $\beta$ increases from 0, the selection strategy of internal neighborhood and external neighborhood starts to work and thus improves the algorithm's performance. But when $\beta$ becomes too large, most of the search efforts are allocated to information exchange, instead of the search within each task, which will in turn deteriorate the performance of the algorithm. Therefore, the setting of $\beta$ should be able to guarantee a good balance between the search within each task and the information exchange between different tasks. The results indicate that the algorithm is more competitive for the neighborhood selection probabilities of 0.1-0.4.
In addition, Figure \ref{beta1-n-FriedmanTestIGD} shows that the average ranking of the algorithm is relatively robust to changes of the internal neighborhood size for a given neighborhood selection probability. Based on the two sub-figures, it can be concluded that the adoption of external neighborhood has more significant effect on the performance of our proposed algorithm, which is consistent with the analysis and conclusion derived in section \ref{s46}.

\section{Conclusion and Future Work} \label{s5}

In this paper, we proposed a multiobjective multitasking evolutionary algorithm based on decomposition with dual neighborhood. 
For each optimization task, an MOP, like the traditional decomposition strategy, is decomposed into a number of single-objective optimization subproblems by using a set of pre-defined weight vectors.
Further, in addition to an internal neighborhood defined by the Euclidean distance between the weight vectors, each subproblem is also associated with subproblems of other tasks, which is called the external neighborhood, using grey relation analysis.
The internal and external neighborhoods of the subproblem are used to explore the correlations and potentially valuable information between different tasks to further improve the efficiency of solving different tasks.
The experimental results demonstrated that our proposed algorithm works better than four state-of-the-art algorithms and a traditional decomposition-based multiobjective evolutionary algorithm.
Our future work will be focused on how to prevent negative transfer of information between tasks and how to explore the valuable information between tasks in the case of many tasks.


%



%
%

\ifCLASSOPTIONcaptionsoff
  \newpage
\fi



\bibliographystyle{IEEEtran}
\bibliography{IEEEabrv,references}

%
%
%

%




\end{CJK*}
\end{document}